\documentclass[10pt,aps,prd,twocolumn,notitlepage,superscriptaddress,a4]{revtex4-2}
\usepackage{amsmath,amssymb,amsfonts}
\usepackage{bbm,bm}
\usepackage[utf8]{inputenc}
\usepackage[letterpaper,top=2cm,bottom=2cm,left=2.5cm,right=2.5cm,marginparwidth=1.75cm]{geometry}
\usepackage{amsmath,amsfonts,amssymb,graphicx,subfigure}
\usepackage{slashed,color}
\usepackage{multirow}
\usepackage{lineno}
\usepackage{verbatim}
\usepackage[svgnames]{xcolor}
\usepackage{hyperref}

\usepackage{hyperref}
\hypersetup{colorlinks=true, linkcolor=blue, citecolor=ForestGreen,
filecolor=magenta, urlcolor=ForestGreen,
  pdftitle={Bigaran and Ruiz [arXiv:2502.07878]},
  pdfauthor={Bigaran and Ruiz},
  pdfsubject={Subject},pdfcreator={Creator},pdfproducer={Producer},
  pdfkeywords={Effective W Approximation}{Muon Colliders}{Next-to-leading power}{Electroweak Parton Distribution Functions}
}
\graphicspath{{Figs/}}

\newcommand{\mgamc}{\texttt{mg5amc}}
\newcommand{\mgFull}{\texttt{MadGraph5\_aMC@NLO}}

\def\fb{{\rm\ fb}}

\newcommand{\invfb}{{\rm ~fb^{-1}}}

\def\GeV{{\rm\ GeV}}
\def\TeV{{\rm\ TeV}}

\definecolor{magenta}{HTML}{FF00FF}
\definecolor{cornflowerblue}{HTML}{6495ED}
\definecolor{turquoise}{HTML}{40E0D0}

\definecolor{darkgreen}{rgb}{0.0, 0.2, 0.13}
\definecolor{darkmagenta}{rgb}{0.55, 0.0, 0.55}
\definecolor{amber}{rgb}{1.0, 0.6, 0.0}

\newcommand{\orcid}[1]{\,\href{https://orcid.org/#1}{\includegraphics[width=9pt]{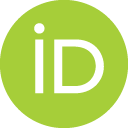}}}
\newcommand{\orcidIB}{0000-0001-6883-953X} 
\newcommand{\orcidRR}{0000-0002-3316-2175} 

\begin{document}
\leftline{}
\rightline{IFJPAN-IV-2025-2, COMETA-2025-1, FERMILAB-PUB-25-0030-T} 

\title{Weak bosons as partons below 10 TeV partonic center-of-momentum}

\author{Innes Bigaran\ \orcid{\orcidIB}} 
\email{innes@vt.edu}
\affiliation{Department of Physics \& Astronomy, Northwestern University, 2145 Sheridan Road, Evanston, IL 60208, USA}
\affiliation{Theoretical Physics Department, Fermilab, P.O. Box 500, Batavia, IL 60510, USA}
\affiliation{Center for Neutrino Physics, Virginia Polytechnic Institute and State University,
Blacksburg, VA 24061, USA}

\author{Richard Ruiz\ \orcid{\orcidRR}}
\email{rruiz@ifj.edu.pl}
\affiliation{Institute of Nuclear Physics -- Polish Academy of Sciences {\rm (IFJ PAN)}, 
Ulica Radzikowskiego, Krak{\'o}w, 31-342, Poland}


\begin{abstract}
We investigate 
the modeling of
 weak boson number densities
 for leptons and hadrons 
 in practical calculations 
in the Standard Model.
In the framework of the 
Effective $W$ Approximation (EWA)
and in the $R_\xi$ and axial gauges,
we derive the unrenormalized, tree-level
parton number densities 
 for weak bosons from massless fermions
at next-to-leading power in the collinear expansion.
Corrections exhibit various pathologies and properties,
including those conjectured but not proven,
and parallel heavy quark factorization.
We suppress pathologies through 
a new set of kinematical consistency conditions.
When satisfied, 
shapes and normalizations of 
full matrix elements for many-leg processes 
can be well approximated by the EWA
and fragmentation contributions at leading power,
suggesting the onset of tree-level factorization.
Findings also suggest that 
the EWA is testable at the LHC 
with $\mathcal{L}=450$ fb$^{-1}$ of
same-sign $WW$ scattering data at $\sqrt{s}=13.6$ TeV.
\end{abstract}

\date{\today} 

\maketitle


\section{Introduction}
Despite confinement in quantum chromodynamics (QCD),
the structure of hadrons can be reliably described 
with the QCD-improved parton model.
A core tenet of this framework is the existence 
of a ``parton sea'' within a hadron, 
i.e., a near continuum of low-energy gluons and
light quark-antiquark pairs.
Importantly, sea partons originate 
from higher energy ``valance'' quarks 
that radiate gluons, 
which subsequently split 
and 
populate lower-energy modes.

In multi-TeV $pp$ collisions,
like those at the Large Hadron Collider (LHC),
 partonic scattering scales 
can be \textit{much} larger than 
the masses of heavy-flavored quarks.
Under such conditions,
the parton sea can be extended to include
perturbatively generated
charm~\cite{Witten:1975bh},
bottom~\cite{Aivazis:1993kh,Aivazis:1993pi},
and top quarks~\cite{Han:2014nja,Dawson:2014pea}.
Since quarks also carry electric charges,
and hence can radiate photons,
the sea distributions in the
proton~\cite{Martin:2004dh,Manohar:2016nzj,Manohar:2017eqh} and
neutron~\cite{Martin:2004dh,Xie:2023qbn} 
also feature a photon component.

In the Standard Model of particle physics (SM), 
quarks $(q)$ and leptons $(\ell)$ carry
weak gauge charges,
and the emergence of 
$W$ and $Z$ bosons 
as (perturbatively generated) sea partons 
in ultra high-energy
hadron collisions
has long been 
predicted~\cite{Dawson:1984gx,Kane:1984bb,Kunszt:1987tk}. 
However, what practically constitutes 
``ultra high energies''  has 
never been firmly established.

In this work, 
we show that the answer is about $800\GeV$.
More specifically, 
when an incoming fermion  radiates 
a (quasi)collinear $W$ with 
at least $E_V=800\GeV$ of energy
($900\GeV$ for $Z$)
in the partonic center-of-momentum (pCM) frame,
the partonic description of weak bosons
becomes a reliable approximation of full matrix elements
at tree level, suggesting the onset of factorization.

Presently, there is no all-orders
 factorization theorem
that extends the parton model to the electroweak (EW) sector.
It is unclear whether such a formulation 
is even  
possible~\cite{Ciafaloni:2000rp,Ciafaloni:2000df,Ciafaloni:2001vt,Ciafaloni:2005fm,Chiu:2007yn,Chiu:2009mg,Manohar:2014vxa,Chen:2016wkt,Bauer:2017isx,Bauer:2018arx,Han:2020uid,Platzer:2022nfu,Frixione:2023gmf}.
Historically, 
predictions for EW boson parton number density functions (PDFs)
 have varied  significantly~\cite{Lindfors:1985yp,Kleiss:1986xp,Kunszt:1987tk,Johnson:1987tj,Abbasabadi:1988ja,Kuss:1995yv,Kuss:1996ww,Accomando:2006mc,Brehmer:2014pka}.
But more recently,
kinematical regimes 
have been proposed where 
predictions for $W/Z$ PDFs might be 
reliable~\cite{Accomando:2006mc,Borel:2012by,Brehmer:2014pka,Costantini:2020stv,Fuks:2020att,Ruiz:2021tdt}.
Here, we derive and firmly establish this regime.

Even at lowest order (LO), 
reliable predictions for weak boson PDFs
can have broad impact. 
LHC data, for example,  
 show that the same-sign $WW$ scattering
signal process, $W_1^\pm W_2^\pm \to W_3^\pm W_4^\pm \to \ell^\pm\ell'^\pm\nu_\ell\nu_{\ell'}$
is well described by LO matrix elements~\cite{CMS:2017fhs,ATLAS:2019cbr,ATLAS:2023sua,CMS:2020etf,Ballestrero:2018anz,BuarqueFranzosi:2021wrv,Dittmaier:2023nac}.
It would be appealing if LO predictions for $W/Z$ PDFs of a proton could
 describe LHC data.
Among applications, 
one could more easily conceptualize and simulate 
 new phenomena~\cite{Henning:2018kys,Bellan:2021dcy,BuarqueFranzosi:2021wrv},
in analogy to heavy quark PDFs~\cite{Dicus:1988cx,Maltoni:2003pn,Han:2014nja,Dawson:2014pea}.
Measuring weak boson PDFs would also 
 constitute laboratory-based probes 
 of the universe’s EW epoch~\cite{Huang:2016cjm,Ramsey-Musolf:2019lsf,OleaRomacho:2022zuz}.
Above all, a complete understanding of 
EW symmetry breaking necessitates 
the determination of weak boson PDFs.

\begin{figure*}[t!]
  \includegraphics[width=\textwidth]{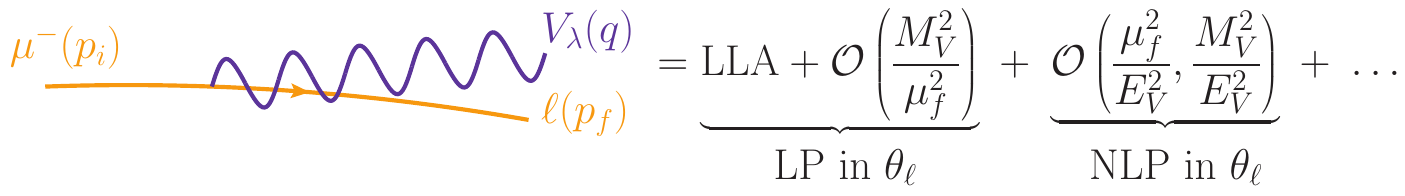}
  \caption{Schematic relationship between the leading log (LLA), leading power (LP), and next-to-leading power (NLP) approximations in collinear $\mu^-\to \ell V_\lambda$ splitting at high energies.
  }
\label{fig:diagram_splitting_NLP}
\end{figure*}

We report a breakthrough in this program.
In the context of the Effective $W$ Approximation (EWA)~\cite{Dawson:1984gx,Kane:1984bb,Kunszt:1987tk}
we present the
unrenormalized, tree-level PDFs for
helicity-polarized weak bosons
from massless, chiral leptons 
at next-to-leading power (NLP) in the collinear expansion.
NLP corrections are process independent,
exhibit phenomenologically 
interesting properties,
and can reconcile reports of 
the EWA's breakdown.
Using these corrections,
we derive a novel 
set of kinematical consistency conditions
based on 
(i) collinear kinematics,
(ii) gauge invariance,
and (iii) positive definiteness.
When enforced,
shapes and normalizations of 
full matrix elements for many-leg processes at tree level
can be well approximated by the EWA
and fragmentation contributions.

Our work continues in the following order:
In Sec.~\ref{sec:setup} we
summarize our theoretical framework
and report our NLP PDFs 
in Sec.~\ref{sec:pdfs}.
In Sec.~\ref{sec:kinematics}
we discuss their 
phenomenological implications 
and in Sec.~\ref{sec:leptons} 
illustrate our findings 
with lepton collisions.
We finish 
with an outlook for  
testing the EWA at the LHC
in Sec.~\ref{sec:conclusion}.
We discuss additional checks and 
technical implementation details 
of our work 
in Appendices \ref{app:usage}, \ref{app:checks}, and \ref{app:scale}.

\section{Setup}\label{sec:setup}
We start by considering the 
inclusive, high-energy,
muon-hadron 
deep-inelastic scattering  process
\begin{align}
 \mu^-(p_i) + h(k) ~\xrightarrow{V_\lambda(q) h\to X}~ 
 \ell(p_f) + X(k')\ .
 \label{eq:disproc}
\end{align}
Here $\ell\in\{\mu^-,\nu_\mu\}$ and
$V\in\{W^\pm,Z\}$ is the exchange boson 
with helicity $\lambda\in\{\pm1,0,S\}$,
momentum $q=p_i-p_f$, 
mass $M_V$,
and
squared virtuality $q^2 \equiv -Q^2 <0$.
The splitting $\mu^-(p_i)\to \ell(p_f) V_\lambda(q)$ 
is illustrated in the left side of 
Fig.~\ref{fig:diagram_splitting_NLP}.
We assume $\sqrt{Q^2}\gtrsim\mathcal{O}(M_V)$
and adopt the following momentum assignments:
\begin{subequations}
\begin{align}
\label{eq:mom_assignments}
p_i^\mu &= E_i\ \left(1,0,0,+1\right),\
 \\
p_f^\mu &=\left((1-z)E_i,\ \vec{p}_T,\ \sqrt{(1-z)^2 E_i^2 -p_T^2} \right)
\\
    &\equiv (1-z)E_i
    \left(1,
    \sin \theta_\ell \cos \phi_\ell,
    \sin \theta_\ell \sin \phi_\ell,
    \cos \theta_\ell \right),
    \nonumber    \\
q^\mu &= p_i^\mu -p_f^\mu \equiv
 \\
\Big(&z E_i,\vert\vec{q}\vert \sin \theta_V \cos \phi_V,
    \vert\vec{q}\vert \sin \theta_V \sin \phi_V,
    \vert\vec{q}\vert \cos \theta_V \Big).
    \nonumber
\end{align}
\end{subequations}
Here, $z=E_V/E_i$ is the fraction of energy $V$ carries from $\mu$. 
From transverse-momentum conservation 
one can derive exact relationships between 
$(\theta_\ell,\phi_\ell)$ and $(\theta_V,\phi_V)$.

To build our PDFs for $V_\lambda$, we use the following basis of polarization vectors in the $R_\xi$ gauge:
\begin{subequations}
\begin{align}
 \varepsilon^\mu (q,\lambda=\pm1) 	&=
 \frac{1}{\sqrt{2}} \Big(0,	-\lambda\cos\theta_V\cos\phi_V + i\sin\phi_V,
 \nonumber\\
-\lambda\cos\theta_V&\sin\phi_V -  i\cos\phi_V,
\lambda\sin\theta_V \Big),
 \\
 \varepsilon^\mu (q,\lambda=0) &=
 \cfrac{E_V}{\sqrt{q^2}\vert\vec{q}\vert } \left( \frac{\vert\vec{q}\vert^2}{E_V}, q_x , q_y , q_z \right)
 \label{eq:pol_0_full}
 \\
 &=
 \frac{q^\mu}{\sqrt{q^2}} + \tilde{\varepsilon}_0^\mu (q), \quad\text{where}
 \\
\tilde{\varepsilon}_0^\mu (q) \equiv& \frac{\sqrt{q^2}}{(E_V+\vert \vec{q}\vert)\vert \vec{q}\vert}\left(-\vert\vec{q}\vert,q_x , q_y , q_z \right)
\label{eq:pol_0_helicty}
\\
 \varepsilon^\mu (q,\lambda=S) &=
 \sqrt{\frac{1}{q^2} + \frac{(\xi-1)}{q^2-\xi M_V^2}}\ q^\mu\ .
\end{align}
\end{subequations}
These obey the completeness relationship
\begin{align}
    &\sum_{\lambda=\pm}\
    \varepsilon_\mu(q,\lambda)\varepsilon^*_\nu(q,\lambda)
    + \varepsilon_\mu(q,0)\varepsilon_\nu(q,0)
    \nonumber\\
    &- \varepsilon_\mu(q,S)\varepsilon_\nu(q,S)
    = -g_{\mu\nu} 
    - \frac{(\xi-1)q_\mu q_\nu}{\left(q^2-\xi M_V^2\right)}\ .
\end{align}

We split the $\lambda=0$  polarization vector
in Eq.~\eqref{eq:pol_0_full}
into its Goldstone $(q^\mu/\sqrt{q^2})$
and gauge $(\tilde{\varepsilon}_0^\mu)$
components.
This makes manifest 
the decoupling of Goldstone modes
since  $q^\mu$
is orthogonal to the $\mu \to \ell$ 
current~\footnote{Note that this decoupling of Goldstone bosons, at least at tree level, 
can possibly soften 
Bloch-Nordsieck violations~\cite{Ciafaloni:2001vt}, i.e., uncancelled infrared EW logarithms.}.
The scalar polarization $(\lambda=S)$ also decouples 
since 
$\varepsilon^\mu(q,S)\propto q^\mu$,
implying results hold for all $\xi$.

We make the usual 
on-shell approximation~\cite{Dawson:1984gx},
and use the polarization vectors 
for external states.
These are obtained by making the replacement $\sqrt{q^2}\to M_V$
in Eqs.~\eqref{eq:pol_0_full}-\eqref{eq:pol_0_helicty},
but nowhere else.
The implications of this have been discussed elsewhere~\cite{Johnson:1987tj,Bernreuther:2015llj,Basu:2025zds}.

\section{Beyond leading log}
\label{sec:pdfs}
We proceed by computing the tree-level helicity amplitudes for
$\mu \to V_\lambda \ell $ splitting.
Next, we expand squared amplitudes to leading power (LP) and
NLP in $\theta_\ell$.
(In contrast,  Refs.~\cite{Altarelli:1987ue,Peskin:1995ev}
evaluate amplitudes \textit{after}  Taylor expanding.)
For $\lambda=0$, only the leading 
$\mathcal{O}(M_V^2/E_V^2)$ term is kept.
We integrate over the phase space of $\ell$.
Because the integral over 
$q^2\sim\mathcal{O}(p_T^2)$  is ultraviolet divergent,
we introduce a regulator $\mu_f>0$ with mass dimension one, 
as in Ref.~\cite{Dawson:1984gx}.
Alternatively, 
one can use dimensional regularization~\cite{Kunszt:1987tk}.

At LP in the $\theta_\ell$ expansion,
the tree-level, helicity-polarized $W/Z$ PDFs
for chiral leptons are 
\begin{subequations}
\label{eq:pdf_xlp}
\begin{align}
    f_{V_+/\ell_L}^{\rm LP}&(z,\mu_f^2) =
    \frac{\tilde{g}^2}{4\pi^2}\
    \frac{g_L^2(1-z)^2}{2z}\
    \nonumber\\
    &
    \left[\log\left(\frac{\mu_f^2+M_V^2}{M_V^2}\right) - \left(\frac{\mu_f^2}{\mu_f^2+M_V^2}\right)\right] ,
    \label{eq:pdf_xlp_vp}\\
    f_{V_-/\ell_L}^{\rm LP}&(z,\mu_f^2) =\frac{\tilde{g}^2}{4 \pi^2}\
    \frac{g_L^2}{2z}\
    \nonumber\\
    &
    \left[\log\left(\frac{\mu_f^2+M_V^2}{M_V^2}\right) - \left(\frac{\mu_f^2}{\mu_f^2+M_V^2}\right)\right] ,
    \label{eq:pdf_xlp_vm}\\
    f_{V_0/\ell_L}^{\rm LP}&(z,\mu_f^2) = 
    \frac{\tilde{g}^2}{4\pi^2}
    \frac{g_L^2  (1-z)}{z}
    \left(\frac{\mu_f^2}{\mu_f^2+M_V^2}\right) ,
    \label{eq:pdf_xlp_v0}\\
    f_{V_\lambda/\ell_R}^{\rm LP}&(z,\mu_f^2) = \left(\frac{g_R}{g_L}\right)^2 \times 
    f_{V_{-\lambda}/\ell_L}^{\rm LP}(z,\mu_f^2) ,
    \label{eq:pdf_xlp_rh}
\end{align}
\end{subequations}
where the couplings are defined in Table~\ref{tab:ewa_coup}.
Keeping the leading term 
when $\mu_f^2\gg M_V^2$ 
recovers the usual PDFs  
in the leading log approximation 
(LLA)~\cite{Dawson:1984gx,Kane:1984bb,Kunszt:1987tk}.
The relationship between collinear expansions 
at LLA, LP, and NLP is illustrated 
in Fig.~\ref{fig:diagram_splitting_NLP}.

At NLP in the collinear expansion, we obtain:
\begin{subequations}
\label{eq:pdf_nlp}
\begin{align}
f_{V_+/\ell_L}^{\rm NLP}(z,\mu_f^2) &=
f_{V_+/\ell_L}^{\rm LP}(z,\mu_f^2)\ \left[1+\frac{(2-z)}{(1-z)}\frac{M_V^2}{E_V^2}\right]\
\nonumber\\
&
-\
\frac{\mu_f^2}{4 E_V^2}\ (2-z)\
f_{V_0/\ell_L}^{\rm LP}(z,\mu_f^2)
\label{eq:pdf_nlp_vp}
\\
f_{V_-/\ell_L}^{\rm NLP}(z,\mu_f^2) &=
f_{V_-/\ell_L}^{\rm LP}(z,\mu_f^2)\ \left[1+(2-z)\frac{M_V^2}{E_V^2}\right]\
\nonumber\\
&
-\
\frac{\mu_f^2}{4 E_V^2}\
\frac{(2-z)}{(1-z)}
f_{V_0/\ell_L}^{\rm LP}(z,\mu_f^2)\ ,
\label{eq:pdf_nlp_vm}
\\
f_{V_0/\ell_L}^{\rm NLP}(z,\mu_f^2) &= f_{V_0/\ell_L}^{\rm LP}(z,\mu_f^2)
\nonumber\\
-\frac{M_V^2}{2 E_V^2}& \left[f_{V_+/\ell_L}^{\rm LP}(z,\mu_f^2)+f_{V_-/\ell_L}^{\rm LP}(z,\mu_f^2)\right]\ ,
\label{eq:pdf_nlp_v0}\\
    f_{V_\lambda/\ell_R}^{\rm NLP}(z,\mu_f^2) &= 
\left(\frac{g_R}{g_L}\right)^2\ \times\ f_{V_{-\lambda}/\ell_L}^{\rm NLP}(z,\mu_f^2)\ .
\end{align}
\end{subequations}
For quarks, weak boson PDFs 
at LLA, LP, and NLP are the same,
up to gauge quantum numbers.

\section{A novel set of constraints}
\label{sec:kinematics}
From Eq.~\eqref{eq:pdf_nlp} 
several observations can be made:

(i) NLP PDFs can be written 
as linear sums of LP PDFs for different  helicities
with mass-over-energy coefficients.
This is suggestive of helicity inversion.

(ii) NLP PDFs for $\lambda=\pm1$
 grow negative with increasing $\mu_f^2/E_V^2$,
 violating positive definiteness of PDFs at LO.
Here,  
$\mu_f/E_V \sim p_T^V/p_z^V\sim\theta_V$, and 
the breakdown occurs because momenta are taken 
outside the collinear limit.
We stress that taking $\mu_f\ll E_V$ is contrary 
to the standard practice of setting 
$\mu_f \sim \mathcal{O}(E_V)$.
However, the practice is only justifiable 
when PDFs are resummed
and $\mu_f$ is a renormalization scale
as in Refs.~\cite{Kunszt:1987tk,Ciafaloni:2005fm,Bauer:2017isx,Han:2020uid}.
%

\begin{table}[!t]
\def\arraystretch{1.5}
\resizebox{\columnwidth}{!}{
\begin{tabular}{|c|c|c|c|}
\hline
Vertex & $\tilde{g}$ & $g_R^f$  & $g_L^f$  \\
\hline
$W-f-f'$	& $\frac{g}{\sqrt{2}}$	& $0$	& $1$	\\
$Z-f-f$	    & $\frac{g}{\cos\theta_W}$	& $-Q^f\sin^2\theta_W$	& $(T_3^f)_L-Q^f\sin^2\theta_W$	 \\
\hline
\end{tabular}
}
\caption{
Coupling definitions for fermions $f,f'$ with weak isospin  $(T_3^f)_L=\pm1/2$ and electric charge $Q^f$.
}
\label{tab:ewa_coup}
\end{table}

(iii) NLP PDFs for $\lambda=0$
 grow negative with decreasing $E_V^2/M_V^2$, 
 reflecting a violation of 
 the $\mathcal{O}(M_V^2/E_V^2)$ expansion.
Again, $z=E_V/E_\ell>M_V/E_\ell$
is a nonstandard restriction 
when applying the EWA.

(iv) In Eq.~\eqref{eq:pdf_nlp},
$\mathcal{O}(M_V^2/E_V^2)$ 
and 
$\mathcal{O}(\mu_f^2/E_V^2)$ terms 
violate Bjorken scaling.
In the deep-inelastic limit,
$E_V$ and $Q^2\sim \mu_f^2 \gtrsim M_V^2$ 
are large, 
$z\propto Q^2/E_V$ is fixed,
$\mathcal{O}(\mu_f^2/E_V^2)$ terms vanish, 
and scaling is recovered.

Corrections to weak boson PDFs 
beyond LP have been estimated 
for several  
processes~\cite{Lindfors:1985yp,Johnson:1987tj,Altarelli:1987ue,Kauffman:1989aq,Abbasabadi:1988ja,Accomando:2006mc,Borel:2012by,Bernreuther:2015llj,Ruiz:2021tdt},
albeit with mixed findings regarding their size.
After careful study of these works,
we can attribute such disagreements 
to inappropriate choices of $z$ and $\mu_f$.
For the photon, fermion-mass 
corrections beyond the LLA are 
well known~\cite{Frixione:1993yw}.
The PDFs in Eq.~\eqref{eq:pdf_nlp} 
have not been previously reported.

\begin{figure*}[t!]
\subfigure[]{\includegraphics[width=\columnwidth]{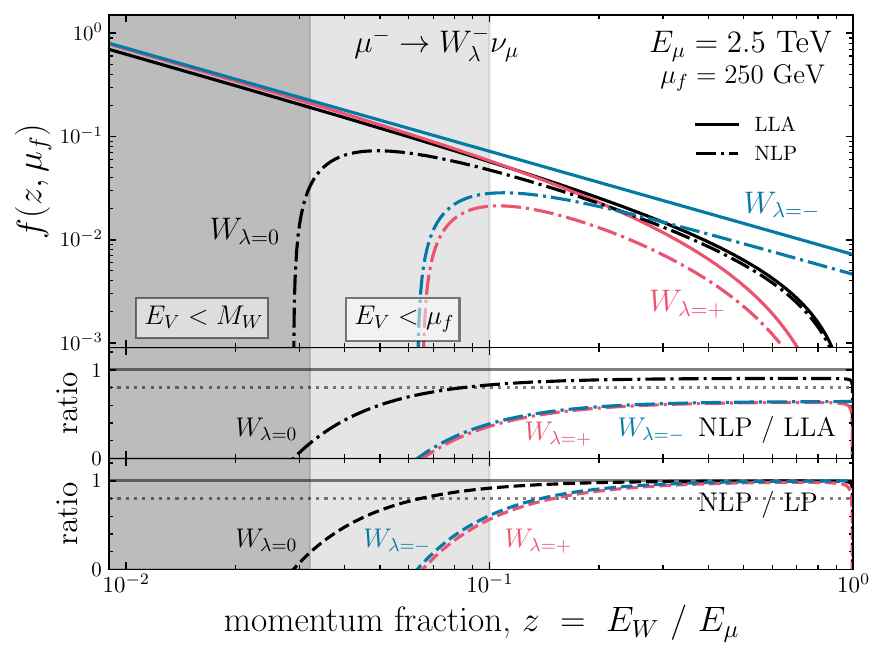}
}
\subfigure[]{\includegraphics[width=\columnwidth]{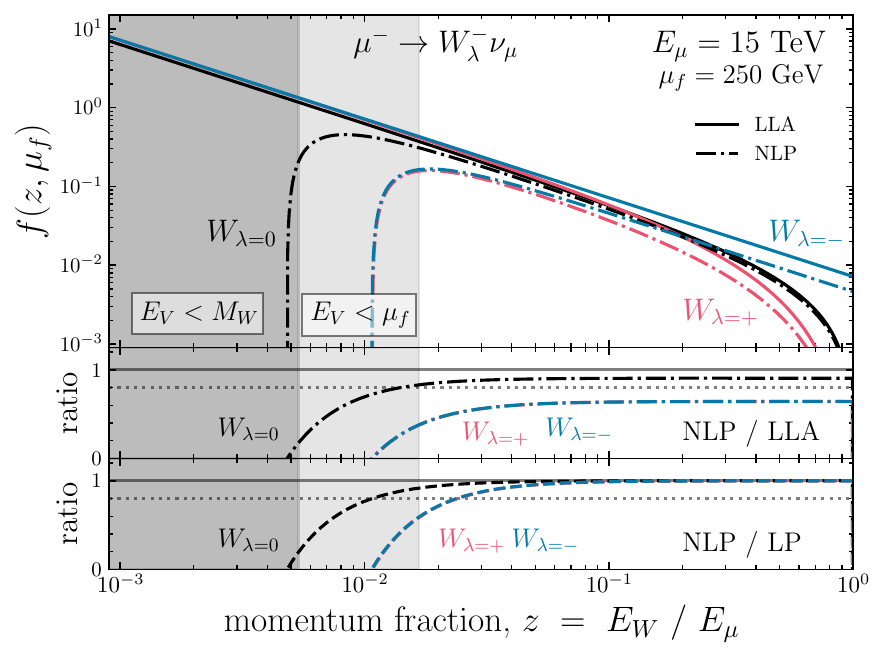}
}
  \caption{Top panel:
  For (a) $E_\mu = 2.5\TeV$ and (b) $E_\mu=15\TeV$,
  the NLP (dash-dot) and LLA (solid) PDFs
  for $W$ in $\mu^-\to W_\lambda^-\nu_\mu$,
  as a function $z=E_W/E_\mu$ with $\mu_f=250\GeV$.
  Middle (Bottom): ratio of NLP and LLA (LP) PDFs.
}
\label{fig:ewaNLP_Wpdf_vs_xx}
\end{figure*}

Beyond the $R_\xi$ gauge,
we check gauge dependence
using the EW axial gauge~\cite{Dams:2004vi}
with reference vectors
\begin{subequations}
\begin{align}
    {\rm lightcone}\ {\rm(LC)}\ &:\ n^\mu\ =\ 
    (1,-\hat{q})\ ,\ n^2=0\ ,
    \\
    {\rm temporal}\ {\rm(TL)}\ &:\ n^\mu\ =\ 
    (1,\ \vec{0})\ ,\ n^2=+1\ ,
    \\
    {\rm spatial}\ {\rm(SL)}\ &:\ n^\mu\ =\ 
    (0,-\hat{q})\ ,
     n^2=-1\ .
\end{align}
\end{subequations}
The polarization vectors for $\lambda=\pm1$ helicities 
are the same here as in the $R_\xi$ gauge.
For $\lambda=S$, we have $\varepsilon^\mu (q,\lambda=S) = 0$
while for $\lambda=0$,
\begin{align}
\label{eq:pol_0_helicty_axial}
\varepsilon^\mu &(q,\lambda=0) = 
\frac{\sqrt{q^2}\left[\frac{ n^2}{(q\cdot n)}q^\mu -n^\mu\right]}{\sqrt{(q\cdot n)^2-n^2 q^2}}\ .
\end{align}
Together, these obey the completeness relationship
\begin{align}
    \sum_{\lambda=\pm}\ &
    \varepsilon_\mu(q,\lambda)\varepsilon^*_\nu(q,\lambda)
    + \varepsilon_\mu(q,0)\varepsilon_\nu(q,0)
    - \varepsilon_\mu(q,S)\varepsilon_\nu(q,S)
    \nonumber\\
    &= -g_{\mu\nu} +
\frac{n_\mu q_\nu + n_\nu q_\mu}{(n\cdot q)} - \frac{n^2}{(q\cdot n)^2}q_\mu q_\nu\  .
\end{align}
In this gauge the 
on-shell approximation~\cite{Dawson:1984gx}
 is obtained by making the replacement
$q^2\to M_V^2$~\cite{Basu:2025zds}.

In the LC gauge, 
the $q^\mu$ term 
in Eq.~\eqref{eq:pol_0_helicty_axial}
vanishes and 
the $n^\mu$ term reduces to 
$\tilde{\varepsilon}_0^\mu(q)$ in Eq.~\eqref{eq:pol_0_helicty}.
Hence, for all three helicities the PDFs at all powers
of $\theta_\ell$
are the same as those in the $R_\xi$ gauge.
In the TL and SL gauges,
we also obtain the same PDFs at LP and LLA.
The robustness of $W/Z$ PDFs at LP
across multiple gauges 
has not been previously reported.

At NLP, only the transverse PDFs are the same
in the TL and SL gauges.
For the longitudinal PDF, 
we obtain $\mathcal{O}(M_V^2/E_V^2)$  
terms 
that differ from Eq.~\eqref{eq:pdf_nlp_v0}.
This gauge dependence is a feature of axial gauges.
Changing $n^\mu$ corresponds to shuffling 
 terms between individual graphs. 
 Indeed, interference terms 
 in the $R_\xi$ and LC gauges  
 differ from those in the TL and SL gauges
   by 
 $\mathcal{O}(M_V^2/E_V^2)$ and 
 $\mathcal{O}(M_V\mu_f/E_V^2)$ terms.

Altogether, EWA predictions 
can be trusted but only when NLP corrections
are negligible, i.e., when 
\begin{align}
 (M_V^2/E_V^2)\ll1\ \text{and}\    (\mu_f^2/E_V^2)\ll1\ 
 \text{(lab\ frame)} .
\label{eq:guidelines}
\end{align}

These kinematical criteria 
are the key finding of our work
and can resolve historic discrepancies between 
EWA and full matrix element computations.

To illustrate their impact 
we show in the top panel of
Fig.~\ref{fig:ewaNLP_Wpdf_vs_xx}
the NLP (dash-dot) and LLA (solid) PDFs
  for $W$ in $\mu^-\to W_\lambda^-\nu_\mu$ splitting
  as a function of energy fraction $z=E_W/E_\mu$,
  at (a) $E_\mu = 2.5\TeV$ 
  and (b) $E_\mu = 15\TeV$ 
  with $\mu_f=250\GeV$.
  Configuration (a) corresponds
  to an LHC-like valance parton at $\sqrt{s}=14$ TeV
  or a benchmark $\mu^+\mu^-$ collider at 
  $\sqrt{s}=5\TeV$~\cite{Black:2022cth,P5:2023wyd,Accettura:2023ked}.
  In the middle (bottom) panel, 
  we show the NLP-to-LLA (NLP-to-LP) PDF ratios.

Pathologies emerge as $z\to0$. 
When $E_W<M_W$, the NLP PDF 
for $\lambda=0$ is negative,
while for $\lambda=\pm1$, 
this occurs when $E_W<\mu_f$.
The bottom panels show that 
pure NLP terms are 
suppressed when $E_W>\mu_f$.
This means that for moderate and large $z$
the NLP-to-LLA ratio
can be interpreted as the LP-to-LLA ratio. 
As $\mu_f=250$,
we are not in the large-log limit, 
and $\mathcal{O}(1)$ terms at LP are 
about $\delta f^{\rm LP}/f^{\rm LLA}\sim -10\%\ (-35\%)$ 
for $\lambda=0\ (\pm1)$ when $z\gtrsim0.1$.

\section{EWA in high-energy lepton collisions}
\label{sec:leptons}
To further illustrate the impact of Eq.~\eqref{eq:guidelines}, 
we consider in $e^+\mu^\mp$ collisions the 
following processes
\begin{align}
\label{eq:wwScattProc}
 W_\lambda^+ W_{\lambda'}^- \to hh,\ 
  \gamma\gamma\gamma,\ 
 \text{and}\
 W_\lambda^+ W_{\lambda'}^+\to W^+ W^+  
\end{align}
at LO and at $\sqrt{s}=5\TeV$.
We focus on these because
(i) $W^+W^-\to hh$ is mediated 
almost exclusively by $W_0W_0$ 
scattering~\cite{BuarqueFranzosi:2019boy,Ruiz:2021tdt}.
(ii) Similarly, $W^+W^-\to \gamma\gamma\gamma$ is  
driven by $W_T^+ W_T^-$ $(T=\pm1)$ 
scattering~\cite{BuarqueFranzosi:2019boy,Ruiz:2021tdt}.
The full $2\to5$ matrix element (ME) also 
contains non-trivial gauge cancellations
and gauge-invariant, bremsstrahlung 
sub-contributions~\cite{Was:2004ig}.
(iii) The full $2\to4$ ME for $W^+W^+$ production
contains large gauge 
cancellations and weak fragmentation contributions 
off initial- and final-state legs.

\begin{figure*}[t!]
\subfigure[]{\includegraphics[width=.32\textwidth]{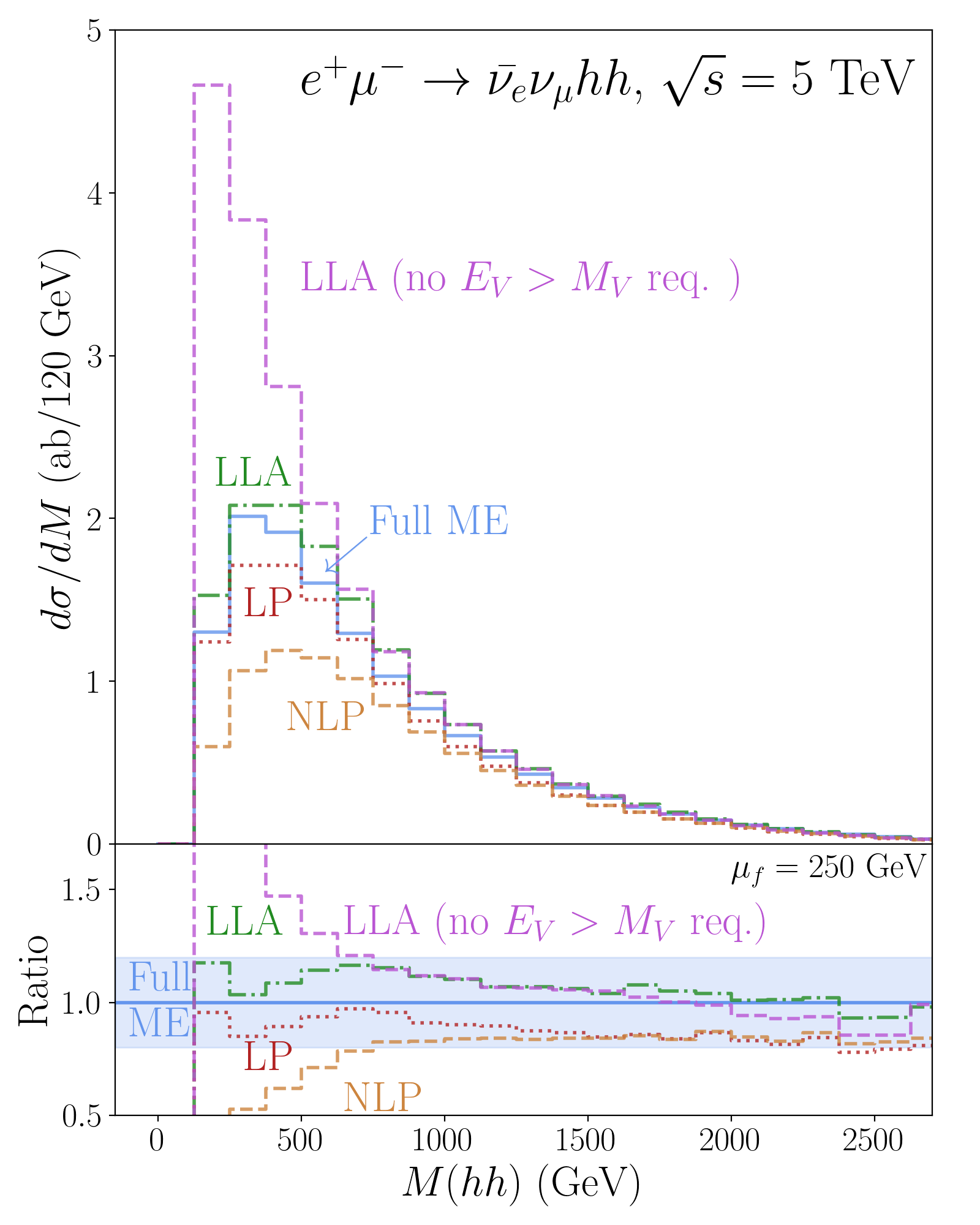}
\label{fig:ewaNLP_Mhh_5TeV}
}
\subfigure[]{\includegraphics[width=.32\textwidth]{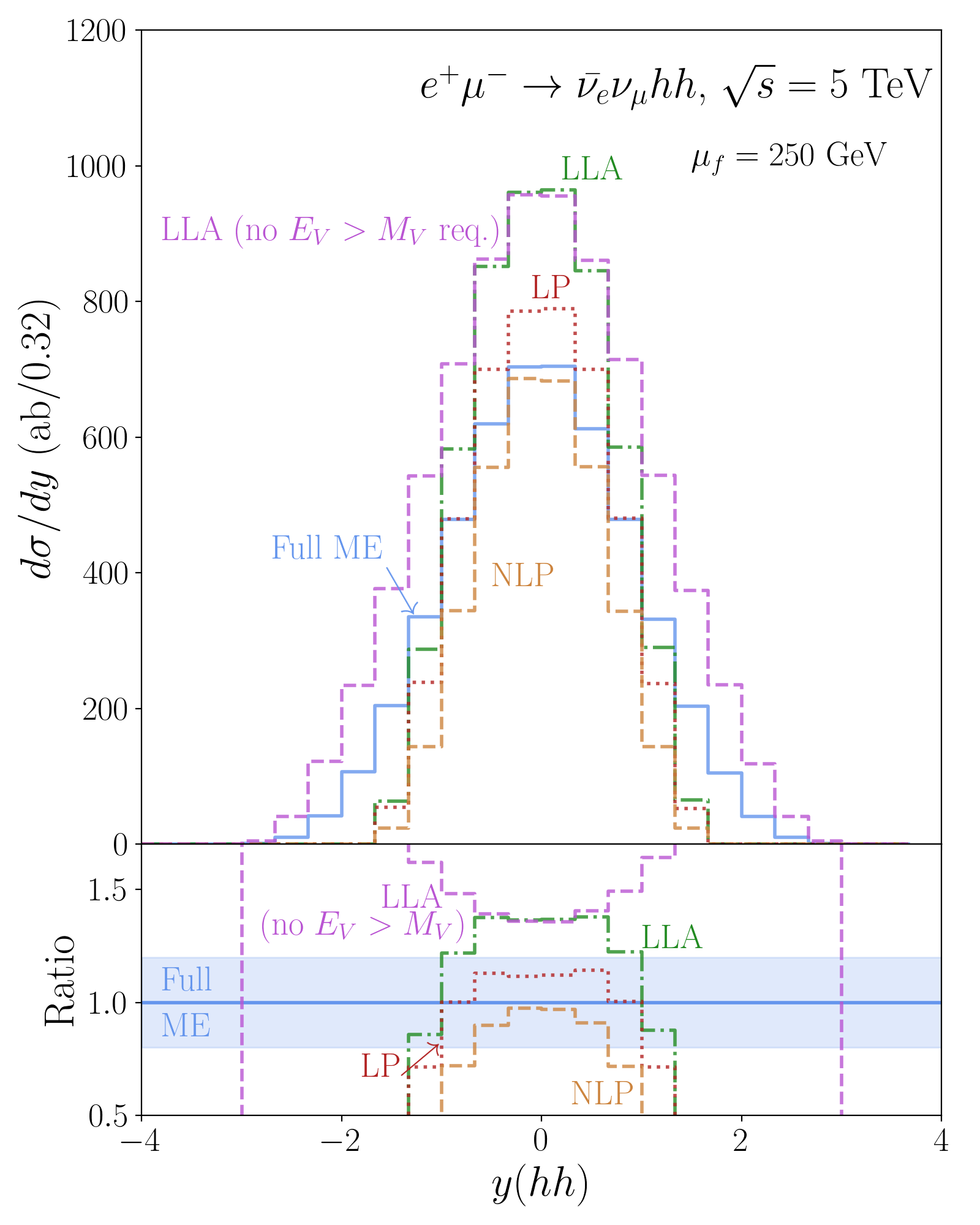}
\label{fig:ewaNLP_yhh_5TeV}
}
\subfigure[]{\includegraphics[width=.32\textwidth]{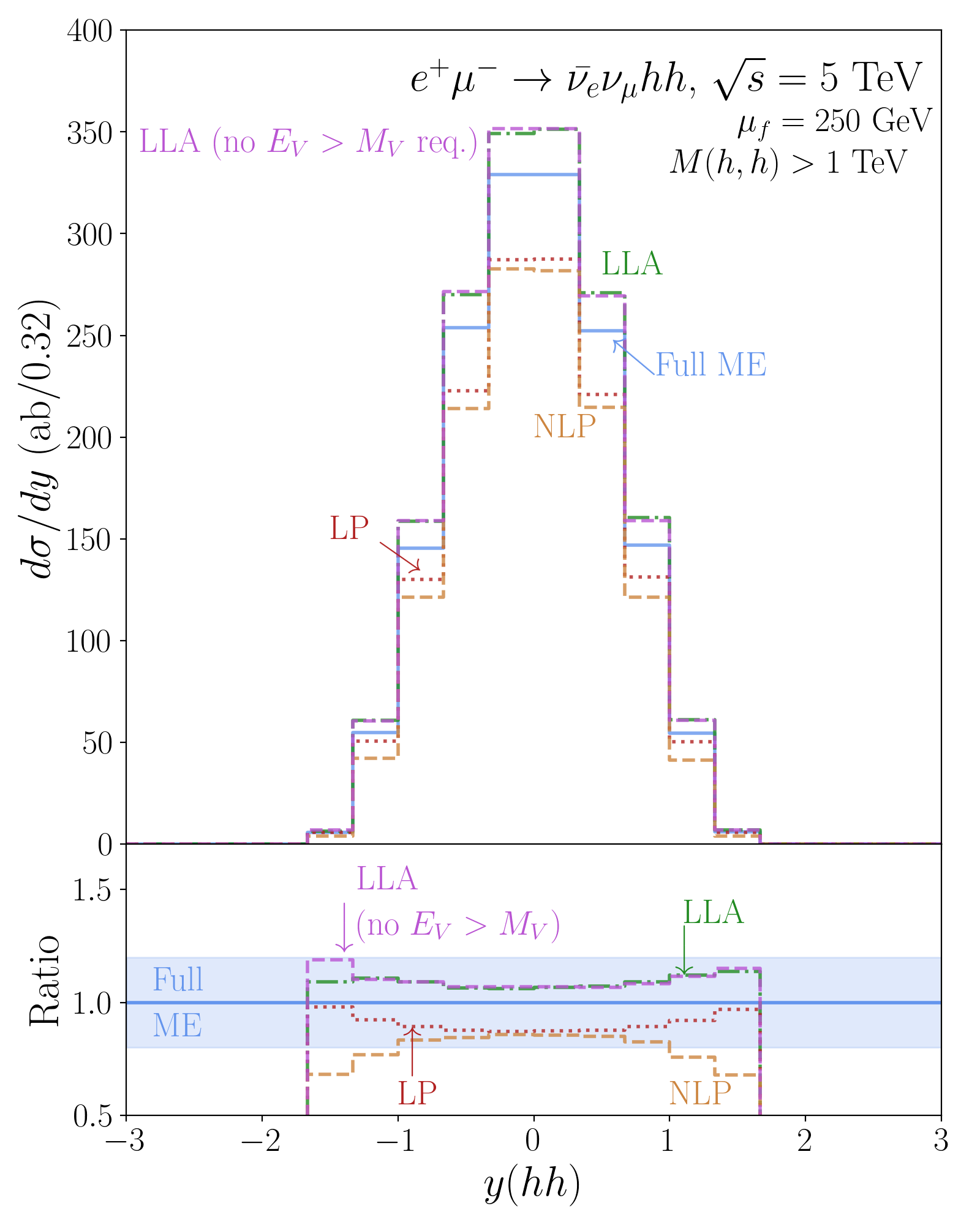}
\label{fig:ewaNLP_yhh_5TeV_cut}
}
\caption{(a) Upper: The invariant mass distribution of the $(hh)$-system 
  in $e^+\mu^-\to \overline{\nu_e}\nu_\mu hh$ at 
  LLA without imposing $E_V>M_V$ (dashed),
  LLA (dot-dashed),
  LP (dotted), and 
  NLP (dashed), 
  along with the full $2\to4$ ME (solid) at $\sqrt{s}=5\TeV$.
  Lower: ratios with respect to the full ME.
  (b,c) Same as (a) but for the rapidity of the $(hh)$-system,
  and for $M(hh)>1\TeV$.
  }
\label{fig:ewaNLP_MWW_HH}
\end{figure*}

To simulate these processes,
we implemented 
the LP and NLP PDFs given in  
Eqs.~\eqref{eq:pdf_xlp} and \eqref{eq:pdf_nlp}
for weak bosons from electrons and muons 
into the simulation 
framework 
\texttt{MadGraph5\_aMC@NLO}
\cite{Stelzer:1994ta,Alwall:2014hca,BuarqueFranzosi:2019boy,Ruiz:2021tdt}.
(New features are available from \texttt{v3.7.0} onward; see also App.~\ref{app:usage} for usage support.)
Total $(\sigma)$ and differential $(d\sigma)$
cross sections in the EWA 
are obtained assuming the factorization formula
\begin{align}
\label{eq:factFormula}
    \sigma_{e^+\mu^-\to \mathcal{F}+X} &= 
    f_{V_\lambda/e^+} 
    \otimes
    f_{V'_{\lambda'}/\mu^-}
    \otimes
    \hat{\sigma}_{V_\lambda V'_{\lambda'}\to \mathcal{F}}\ .
\end{align}
Here, $\hat{\sigma}$ is the ``partonic'' cross section 
for the processes in Eq.~\eqref{eq:wwScattProc}. 
 $\otimes$ denotes a convolution over $z_i$
for incoming boson $i$.
Unless noted, we require 
$z_{1(2)}>z_{\rm min}=(M_W/E_{e(\mu)})$. 
Initial-state $W$'s are 
collinear to the beam $(\vert\vec{q}_{Ti}\vert=0)$ and
on-shell, with $M_W^2\neq0$.
Helicities and energies  
are defined 
in the frame of the $(WW)$-system.
We set $\mu_f = 250\ (450)\ [340]\ \GeV$ for 
$hh\ (\gamma\gamma\gamma)\ [W^+W^+]$ production, 
which approximates the threshold scale
for each process.
For non-EWA MEs,
 all interfering diagrams are considered.

To regulate infrared poles
in full and approximated MEs, 
we impose on $V\in\{\gamma,W\}$
\begin{align}
\label{eq:cuts_photons}
p_T^V > 150\GeV,\ \vert\eta^V\vert <3,\ \text{and}\
\Delta R_{\gamma\gamma} > 0.4\ .
\end{align}
Since $\sqrt{s}=5\TeV$, Eq.~\eqref{eq:cuts_photons} is 
sufficient to ensure  MEs are 
well defined/perturbative in the CSS sense~\cite{Collins:1984kg}.
However, 
this is not guaranteed for larger $\sqrt{s}$
or for alternative multi-leg processes~\cite{Han:2020uid}.

In the upper panel of Fig.~\ref{fig:ewaNLP_Mhh_5TeV} we show the
invariant mass distribution of the $(hh)$-system 
in $W^+W^-\to hh$ as predicted by the EWA
at LLA (dot-dashed), LP (dotted), and NLP (dashed),
  as well as by the full $2\to4$ ME
  for $e^+\mu^-\to \overline{\nu_e}\nu_\mu hh$   (solid).
  The lower panel shows the ratios with respect to the full ME
  with a {$\pm20\%$} tolerance band.

Over the entire $M(hh)=M(WW)$ mass range, 
disagreements between the full ME, LLA, and LP distributions
fall within {$\mathcal{O}(\pm20\%)$}.
For low $M(WW)$, 
NLP predictions are far below the full ME. 
However, NLP and LP predictions converge 
for $M(WW)\gtrsim 1.6\TeV$,
which corresponds to $E_W\sim 800\GeV$ 
and  $(M_W^2/E_W^2) \sim \mathcal{O}(1\%)$.
For comparison, we also plot the LLA prediction 
allowing $z<M_V/E_\ell$ (dashed),
which leads to the EWA 
overestimating the low $M(WW)$ region.
We find similar agreement 
in the rapidity of the $(hh)$ system
when omitting [Fig.~\ref{fig:ewaNLP_yhh_5TeV}]
or applying [Fig.~\ref{fig:ewaNLP_yhh_5TeV_cut}]
the criteria of Eq.~\eqref{eq:guidelines}.

For $M(hh)>2m_h\ (1.6\TeV)$, 
cross sections [fb] with 3-point scale uncertainties [\%] 
are listed in Table~\ref{tab:3point}.
Also shown are the deviations from the full ME [\%].
The LLA best recovers the full ME at large $M(WW)$ 
but with an 
untrustworthy uncertainty of $\mathcal{O}(1\%)$ 
since $\mathcal{O}(M_W^2/\mu_f^2)$ terms 
are neglected.
LP and NLP predictions feature slightly worse agreements 
but more reliable uncertainties.
Interestingly, setting $\mu_f=4m_h\sim 500\GeV$ (not shown),
which is roughly at the maximum of the $M(WW)$ distribution,
reconciles (N)LP and full ME predictions.

\begin{table}[!t]
    \centering
  \resizebox{.9\columnwidth}{!}{  \begin{tabular}{c| r l |c}
 $\sigma$ [fb]\ & $M(hh)>2m_h$ & $(1.6\TeV)$ & $(\sigma^{\rm EWA}-\sigma^{\rm Full})/\sigma^{\rm Full}$ \\
\hline
  Full
  & $1.66$ & $(0.186)$
  & -- (--)
  \\
  LLA
  & $1.86 ^{+3\%}_{-1\%}$ &
  $\left(0.194^{+1\%}_{<-0.5\%}\right)$
  & $+12\%\ (+4\%)$
   \\
   LP
   & $1.51 ^{+18\%}_{-40\%}$ &
   $\left(0.158  ^{+16\%}_{-39\%}\ \right)$
   & $-9\%\ (-14\%)$
   \\
   NLP
   & $1.16 ^{+0\%}_{-29\%}$ &
   $\left(0.157 ^{+15\%}_{-39\%}\right)$
   & $-30\%\ (-16\%)$\\
\end{tabular}}
    \caption{
    Predicted cross sections $(\sigma)$ [fb] for 
    $e^+\mu^-\to \overline{\nu_e}\nu_\mu hh$ 
    at $\sqrt{s}=5\TeV$
    using the full ME (row 1) 
    and the EWA at various accuracies (rows 2-4),
    along with standard 3-point scale uncertainties,
    for $M(hh)>2m_h\ (1.6\TeV)$.
    Also shown (column 3) are the deviations [\%] 
    from the full ME.}
    \label{tab:3point}
\end{table}

To model the $\gamma\gamma\gamma$ process,
we consider the incoherent sum of 
the EWA channel $(\sigma_{3\gamma}^{\rm EWA})$ 
and a bremsstrahlung term $(\sigma_{3\gamma}^{\rm brem})$.
The $\sigma_{3\gamma}^{\rm brem}$ contribution 
is estimated by scaling the rate for the 
$e^+\mu^-\to \overline{\nu_e}\nu_\mu \gamma\gamma$ process 
by a QED Sudakov factor: 
\begin{subequations}
\begin{align}
\sigma_{3\gamma}^{\rm brem.} &= \mathcal{S}_{\rm QED}\ \times\ \sigma^{\rm Full}(e^+\mu^-\to \overline{\nu_e}\nu_\mu \gamma\gamma) , \\
\label{eq:sudakov_qed}
    \mathcal{S}_{\rm QED} &= \frac{\alpha}{4\pi}\ \log^2\ \left(\frac{\mu_S^2}{(p_{T,\min})^2}\right)\ ,
\end{align}
\end{subequations}
with $p_{T,\min}=150\GeV$ and $\mu_S=\mu_f$.
For each EWA contribution, 
we then rescale $\sigma_{3\gamma}^{\rm EWA}$ to 
$\sigma_{3\gamma}^{\rm EWA+brem}=\sigma_{3\gamma}^{\rm EWA}+\sigma_{3\gamma}^{\rm brem}$.
Numerically,
$\sigma_{3\gamma}^{\rm EWA}$ and $\sigma_{3\gamma}^{\rm brem}$ are comparable.
Due to $\mathcal{O}(\mu_f^2/E_V^2)$ corrections,
$\sigma_{3\gamma}^{\rm NLP}$ goes negative 
for $M(\gamma\gamma\gamma) \lesssim 1250\GeV$.

\begin{figure}[t!]
\centering 
\includegraphics[width=.64\columnwidth]{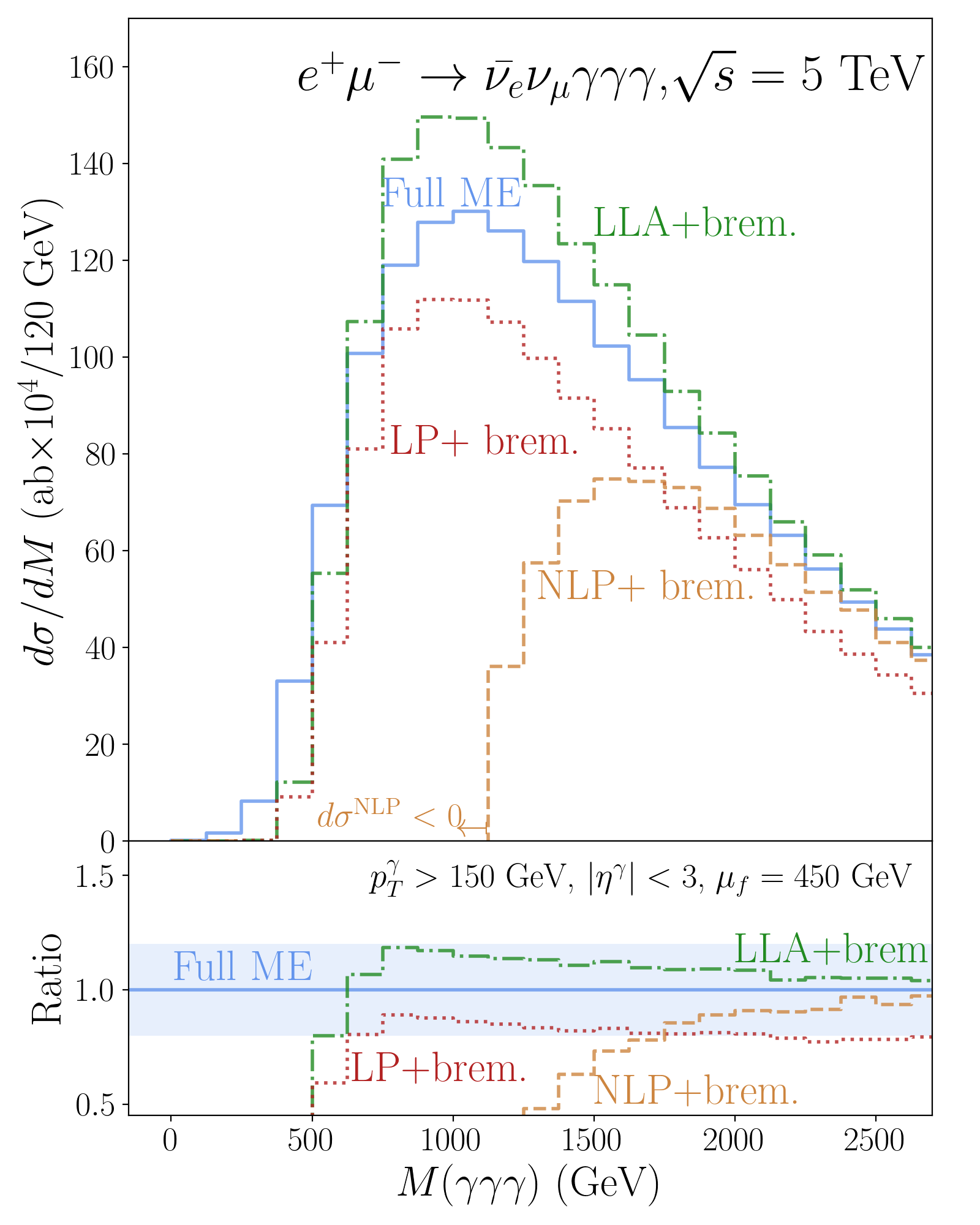}
\caption{Same as Fig.~\ref{fig:ewaNLP_Mhh_5TeV} but
  for the $(3\gamma)$-system in the  $e^+\mu^-\to \overline{\nu_e}\nu_\mu 3\gamma$ process,
  including EWA and bremsstrahlung contributions.
  }
\label{fig:ewaNLP_MWW_AAA_5TeV_Sud}
\end{figure}

In Fig.~\ref{fig:ewaNLP_MWW_AAA_5TeV_Sud} we show the invariant mass
of the $(\gamma\gamma\gamma)$-system 
for the full ME 
and EWA+brem.\ combinations.
Remarkably, we obtain good agreement 
despite the na\"ive modeling of $\sigma_{3\gamma}^{\rm brem.}$.
Predictions for LLA+brem.\ and LP+brem.\ fall 
within {$\mathcal{O}(20\%)$ of the full ME
for $M(3\gamma)\gtrsim 800\GeV$.}
The NLP+brem.\ curve achieves this 
for {$M(WW)\gtrsim1.8\TeV$} and improves with larger $M(WW)$.

For the $W^+W^+$ process, we again combine 
the EWA channel $(\sigma_{WW}^{\rm EWA})$ 
with a fragmentation term $(\sigma_{WW}^{\rm brem})$.
We estimate $\sigma_{WW}^{\rm brem}$ by scaling the 
$e^+\mu^+\to \overline{\nu_{\ell'}}\ell^+ W^+$  process
by the Sudakov factor in Eq.~\eqref{eq:sudakov_qed} 
 but with the replacements 
 $\alpha\mapsto\alpha_W$ and 
 $p_{T,\min}^2\mapsto 
 p_{T,\min}^2+M_W^2$~\cite{Bauer:2016kkv}.
We approximate the momentum of $W$ in final-state 
 $\ell\to W\nu$ splitting by $p_{W}\approx p_\ell/2$
and adapt Eq.~\eqref{eq:cuts_photons} 
for $\ell$ ($p_T^\ell>300\GeV$, etc.).

\begin{figure}[t!]
\centering 
\includegraphics[width=.64\columnwidth]{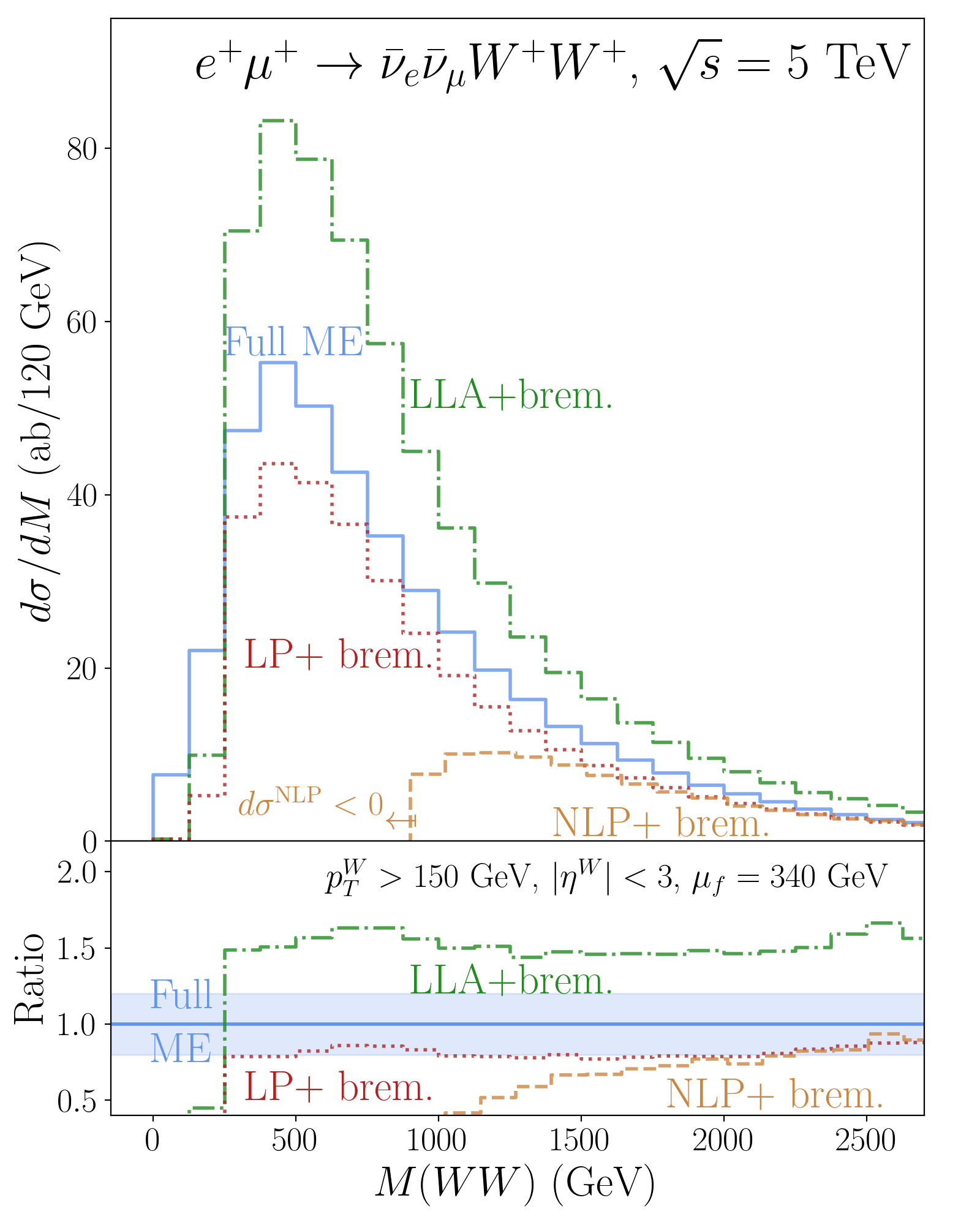}
\caption{Same as Fig.~\ref{fig:ewaNLP_Mhh_5TeV} but
  for the $(WW)$-system in the process $e^+\mu^+\to \overline{\nu_e}\overline{\nu_\mu}W^+W^+$,
  including EWA and bremsstrahlung contributions.
  }
\label{fig:ewaNLP_MWW_WW_5TeV_Sud}
\end{figure}

In Fig.~\ref{fig:ewaNLP_MWW_WW_5TeV_Sud}
we show the invariant mass  
of the $(WW)$-system for the full ME 
and EWA+fragmentation.
LP and NLP curves again largely agree 
with the full ME.
The LLA prediction, however, is 
$\mathcal{O}(50\%)$ too big
due to missing $\mathcal{O}(M_W^2/\mu_f^2)$ terms.

When using renormalized collinear PDFs, 
the scale invariance of 
physical observables guarantees 
the existence of a (soft) Sudakov factor~\cite{Contopanagos:1996nh}.
Assuming the formula 
of Eq.~\eqref{eq:factFormula} holds, 
we 
anticipate better agreement 
between full MEs and 
predictions based on EWA and fragmentation
when parton-shower, matching, 
and  resummation methods are 
employed~\cite{Brooks:2021kji,Ruiz:2021tdt,Bredt:2022dmm,Denner:2024yut,Ma:2024ayr}.
Even at LO, the impact can be significant~\cite{Ahrens:2009cxz,Becher:2014oda},
but demonstrating this 
requires investigation.

\section{Conclusion}\label{sec:conclusion}
We report the first NLP-accurate calculation of 
 $W/Z$ PDFs for high-energy fermions [Eq.~\eqref{eq:pdf_nlp}].
 From these PDFs,
 we have derived kinematical criteria 
 [Eq.~\eqref{eq:guidelines}]
 that lead to good agreement between 
 full and approximated MEs,
 including at the differential level,
 and suggest the onset of EW factorization 
 at tree level.
 Partonic behavior emerges when 
weak bosons carry at least 
$E_V=10\times M_V$ of energy
(pCM frame) and $p_T\sim \mu_f\ll E_V$
is near threshold.
This corresponds to 
$\mathcal{O}(M_V^2/E_V^2)\sim1\%$ 
power corrections.

Given this energy scale,
it may be possible to observe the emergence of 
$W$ bosons as partons 
of a proton at the LHC.
At $\sqrt{s}=13.6\TeV$, the same-sign $WW$ scattering
process $pp\to W^\pm W^\pm j j$ 
has a LO cross section
of $\sigma_{\rm LHC13.6}\sim 5.5\ (1.0)\fb$
for $M(WW) > 1\ (1.6)\TeV$ and $M(jj)>500\GeV$.
Taking the $W\to e/\mu$ branching rate BR$^2 \approx 4/81$,
and assuming  
selection and acceptance rates of
$\epsilon\times\mathcal{A}\approx25\%$~\cite{CMS:2017fhs,ATLAS:2019cbr,ATLAS:2023sua,CMS:2020etf},
then about $N=\mathcal{L}\sigma{\rm BR}^2\epsilon\mathcal{A}\approx 30\ (6)$ events are anticipated
with $\mathcal{L}=450\invfb$,
or $N=300\ (60)$ with $\mathcal{L}=4500\invfb$.
This is large enough to be interesting
and motivates further study.
Of particular importance is an 
EWA-compatible parton shower, which is now 
the last missing ingredient for 
simulating 
fully exclusive events at colliders.

\section*{Acknowledgments}
The authors thank
A.\ Apyan,
A.\ Denner,
T.\ Han,
F.\ Olness,
C.\ Quigg,
J.\ Reuter,
G.\ Pelliccioli,
and
the {\mgamc} team
for discussions.
This manuscript has been authored in part by  Fermi Forward Discovery Group, LLC under Contract No. 89243024CSC000002 with the U.S. Department of Energy, Office of Science, Office of High Energy Physics. The work of I.B. was performed in part at the Aspen Center for Physics, supported by a grant from the Alfred P. Sloan Foundation (G-2024-22395). This work was supported by the U.S. Department of Energy under award DE-SC0020262.
R.R. acknowledges the support of Narodowe Centrum Nauki under Grant No. 2023/49/B/ST2/04330 (SNAIL). The authors acknowledge support from the COMETA COST Action CA22130.


\appendix

\section{Usage in {MG5\_aMC@NLO}}
\label{app:usage}

\begin{table}
\begin{tabular}{|c|c|c|}
\hline
\texttt{evaorder} value [\texttt{int}] & PDF accuracy & Eqs.\\
\hline
\texttt{0} & LLA (default) & Ref.~\cite{Ruiz:2021tdt}\\
\texttt{1} & LP & Eq.~\eqref{eq:pdf_xlp}\\
\texttt{2} & NLP & Eq.~\eqref{eq:pdf_nlp}\\
\hline
\end{tabular}
\caption{\texttt{evaorder} settings for {\mgFull}}
\label{tab:MG5}
\end{table}

Calling PDFs at LP and NLP
for helicity-polarized $W$ and $Z$ bosons in
electrons and muons is a new feature in
the simulation framework 
{\mgFull} (v3.7.0 and above)~\cite{Stelzer:1994ta,Alwall:2014hca}.
The development builds on releases for helicity-polarized parton
scattering~\cite{BuarqueFranzosi:2019boy}
and the implementation of the EVA at LLA~\cite{Ruiz:2021tdt}. Sample notebooks and run scripts may be found at Ref.~\cite{EWA_MG5}.

To access LP and NLP PDFs,
nearly identical syntax can be used
as introduced in Ref.~\cite{Ruiz:2021tdt}.
The additions is the \texttt{run\_card} options 
are \texttt{evaorder} and \texttt{eva\_xcut}.
The available values for these two options and their impact 
are found in Tables~\ref{tab:MG5} and \ref{tab:MG5_xmin}.

NLP corrections are only available for weak bosons,
not for the photon (since $m_\gamma=0)$. 
LP corrections to PDFs for photons from leptons
are available as developed in Ref.~\cite{Frixione:1993yw}.
The option \texttt{ievo\_eva}
for setting the EWA evolution variable, 
as developed in Ref.~\cite{Ruiz:2021tdt},
is available only for LLA PDFs.

The  syntax for simulating
the $W^+_0W^-_0\to hh$ process
with the EWA at LP,
as done in Fig.~\ref{fig:ewaNLP_MWW_HH} is 
\begin{verbatim}
set group_subprocesses false
generate w+{0} w-{0} > h h QED=2 QCD=0
output
launch
analysis=off
set ebeam 10000 # GeV
set nevents 10k
set lpp1  -3 # e+ beam
set lpp2   4 # mu- beam
set pdlabel eva
set evaorder 1 # 0=LLA (default), 1=LP, 2=NLP
set eva_xcut 1 # x > MV/Ebeam, 0 = no cut
set fixed_ren_scale true
set fixed_fac_scale1 true
set fixed_fac_scale2 true
set dsqrt_q2fact1 250
set dsqrt_q2fact2 250
set scalefact 1.0
set no_parton_cut
set use_syst true
done
\end{verbatim}
For descriptions of this syntax,
see Refs.~\cite{Alwall:2014hca,BuarqueFranzosi:2019boy,Ruiz:2021tdt}.

\begin{table*}
\begin{tabular}{|c|c l|}
\hline
\texttt{eva\_xcut} value [\texttt{int}] & \multicolumn{2}{c|}{Impact on PDF}\\
\hline
\texttt{1} & $f_{V/\ell}(z,\mu_f^2) = 0$ 
& for $z<M_V/E_\ell$ (lab frame)\\
\texttt{0} & $f^{\rm LLA,\ LP,\ NLP}_{V/\ell}(z,\mu_f^2) = $ 
Ref.~\cite{Ruiz:2021tdt}, Eq.~\eqref{eq:pdf_xlp}, Eq.~\eqref{eq:pdf_nlp} 
& for $z<M_V/E_\ell$ (lab frame)\\
\hline
\end{tabular}
\caption{\texttt{eva\_xcut} settings for {\mgFull}}
\label{tab:MG5_xmin}
\end{table*}

\section{Checks against {MG5\_aMC@NLO}}
\label{app:checks}

\begin{table*}[!t]
\begin{center}
\begin{tabular}{|c|c|c|c|c|}
\hline
Helicity configuration\ $(\lambda_A,\lambda_B)$\ & 
$\hat{\sigma}$ [pb] (no PDFs)\ &
$\sigma^{\rm LLA}$  [pb]\ & 
$\sigma^{\rm LP}$  [pb]\ & 
$\sigma^{\rm NLP}$  [pb]\ \\ 
\hline 
$(+,+)$ & $1.693\times10^{-3}$ & $119.6\times10^{-9}$\ & $41.76\times10^{-9}$\ & $13.56\times10^{-9}$\\
$(+,-)$ & $150.2\times10^{-3}$ & $2960\times10^{-9}$\ &  $1033\times10^{-9}$\ & $581.2\times10^{-9}$\\
$(+,0)$ & $433.3\times10^{-6}$ & $41.5\times10^{-9}$\ & $17.34\times10^{-9}$\ & $11.13\times10^{-9}$\\
$(-,+)$ & $150.2\times10^{-3}$ & $1351\times10^{-9}$\ & $471.7\times10^{-9}$\ & $236.8\times10^{-9}$\\
$(-,-)$ & $1.693\times10^{-3}$ & $119.6\times10^{-9}$\ & $41.75\times10^{-9}$\ & $13.56\times10^{-9}$\\
$(-,0)$ & $433.3\times10^{-6}$ & $31.43\times10^{-9}$\ & $13.14\times10^{-9}$\ & $7.715\times10^{-9}$\\
$(0,+)$ & $433.3\times10^{-6}$ & $31.46\times10^{-9}$\ & $13.15\times10^{-9}$\ & $7.735\times10^{-9}$\\
$(0,-)$ & $433.3\times10^{-6}$ & $41.44\times10^{-9}$\ & $17.32\times10^{-9}$\ & $11.11\times10^{-9}$\\
$(0,0)$ & $50.05$ & $1825\times10^{-6}$ & $913.0\times10^{-6}$ & $817.8\times10^{-6}$\\
\hline
\end{tabular}
\caption{Helicity-polarized partonic (column 2) and leptonic (columns 3-5) cross sections [pb]
for the $W^+(\lambda_A)W^-(\lambda_B)\to hh$ sub-process in $\mu^+\mu^-$ collisions at $\sqrt{s}=5\TeV$,
assuming inputs of Eq.~\eqref{eq:app_inputs}, for various $W^+W^-$ helicity configurations (column 1)
and LLA/LP/NLP PDFs (column 3/4/5). Partonic cross sections (column 2) 
use a hard scattering scale of $Q=1\TeV$.
}
\label{tab:app_checks}
\end{center}
\end{table*}

As a check of our implementation of LP [Eq.~\eqref{eq:pdf_xlp}] and NLP [Eq.~\eqref{eq:pdf_nlp}]
PDFs into {\mgamc}, we independently computed  
the $W^+W^-$ scattering process
\begin{align}
    W^+(p_A,\lambda_A)\ W^-(p_B,\lambda_B)\ \to\ h(p_1)\ h(p_2)\ .
    \label{eq:app_ww_to_hh}
\end{align}
In the Unitary gauge,  the matrix element is  
\begin{align}
    -i\mathcal{M}(\lambda_A,\lambda_B)\ =\ \sum_{d\in\{s,t,u,4\}}\ 
    -i\mathcal{M}_k(\lambda_A,\lambda_B) \ ,
\end{align}
where the sum runs over four  sub-amplitudes.
For helicity $\{\lambda_k\}$ 
and momenta $\{p_k\}$,
the amplitudes are 
\begin{subequations}
    \begin{align}
-i\mathcal{M}_s &= +i\frac{3g^2 m_h^2}{2} 
\frac{\left[\varepsilon(p_A,\lambda_A)\cdot \varepsilon(p_B,\lambda_B)\right]}{Q^2 - m_h^2}\ , 
\\
-i\mathcal{M}_4 &= +i\frac{g^2 }{2}  
\left[\varepsilon(p_A,\lambda_A)\cdot \varepsilon(p_B,\lambda_B)\right]\ ,
\\
-i\mathcal{M}_t &= \frac{+ig^2 }{t-M_W^2}
\Big\{
M_W^2
\left[\varepsilon(p_A,\lambda_A)\cdot \varepsilon(p_B,\lambda_B)\right]
\nonumber\\
&\ + 
\left[\varepsilon(p_A,\lambda_A)\cdot p_1\right]
\left[\varepsilon(p_B,\lambda_B)\cdot p_2\right]
\Big\}\ ,
\\
-i\mathcal{M}_t &= \frac{+ig^2 }{u-M_W^2}
\Big\{
M_W^2
\left[\varepsilon(p_A,\lambda_A)\cdot \varepsilon(p_B,\lambda_B)\right]
\nonumber\\
&\ + 
\left[\varepsilon(p_A,\lambda_A)\cdot p_2\right]
\left[\varepsilon(p_B,\lambda_B)\cdot p_1\right]
\Big\}\ .
    \end{align}
\end{subequations}

In the partonic center-of-mass frame, momenta are
\begin{subequations}
\begin{align}
    p_A &= \frac{Q}{2}\ (1,0,0,+\beta_W)\ ,\ 
    \vert \vec{p}_A\vert = \vert \vec{p}_B\vert = \beta_W \frac{Q}{2}\ ,\\
    p_B &= \frac{Q}{2}\ (1,0,0,-\beta_W)\  , \\
    p_1 &= \frac{Q}{2}\ (1,\beta_h\sin\theta\sin\phi,\beta_h\sin\theta\sin\phi,\beta_h\cos\theta)\ , \\
    p_2 &= \frac{Q}{2}\ (1,-\vec{p}_1),\ 
    \vert \vec{p}_1\vert = \vert \vec{p}_2\vert = \beta_h \frac{Q}{2}\ .
\end{align}    
\end{subequations}
These assignments generate the following contractions
among polarization vectors (all others vanish):
\begin{subequations}
    \begin{align}
        \varepsilon(p_A,\pm1)\cdot \varepsilon(p_B,\pm1) &= -1\ , \\
        \varepsilon(p_A,0)\cdot \varepsilon(p_B,0) &=  -1 + \frac{Q^2}{2M_W^2}\ . 
    \end{align}
\end{subequations}
Similarly, these assignments give the following contractions 
among polarization vectors and momenta:
\begin{subequations}
\begin{align}
    \varepsilon(p_A,\pm1)\cdot p_1 &= \pm \frac{e^{\pm i\phi}}{2\sqrt{2}}Q
    \sqrt{1 - 4r_h}\ \sin\theta\ ,
    \\
    \varepsilon(p_A,0)\cdot p_1 &= \frac{Q^2}{4M_W} \Big[\sqrt{1-4r_W}
    -\sqrt{1-4r_h}\cos\theta\Big]\ ,
    \\
    \varepsilon(p_A,\pm1)\cdot p_2 &= - \left[\varepsilon(p_A,\pm1)\cdot p_1 \right]\ ,
    \\
    \varepsilon(p_A,0)\cdot p_2 &= \frac{Q^2}{4M_W} \Big[\sqrt{1-4r_W}
    +\sqrt{1-4r_h}\cos\theta\Big] \ ,
    \\
    r_W &= \frac{M_W^2}{Q^2}, \quad r_h = \frac{m_h^2}{Q^2}\ .
\end{align}
\end{subequations}

The \textit{helicity-polarized}, parton-level
cross sections is determined (numerically) from 
the formula~\cite{BuarqueFranzosi:2019boy}
\begin{subequations}
\begin{align}
    \hat{\sigma}(\lambda_A,\lambda_B) &= \int dPS_2\ \frac{d\hat{\sigma}(\lambda_A,\lambda_B)}{dPS_2}\ ,\ \text{where}
    \\
   \frac{d\hat{\sigma}(\lambda_A,\lambda_B)}{dPS_2}
    & = \frac{1}{2Q^2\sqrt{1-4r_W}} \vert \mathcal{M}(\lambda_A,\lambda_B)\vert^2,\ 
    \\
    dPS_2 &= \frac{d\cos\theta d\phi}{2(4\pi)^2}\sqrt{1-4r_h}\ .
\end{align}
\end{subequations}

Polarized cross sections at the level of muons 
are obtained through the usual PDF convolutions~\cite{Dawson:1984gx}:
\begin{align}
    \sigma_{\lambda_A\lambda_B}&(\mu^+\mu^-\to hh+X) = 
    f_{W^+_{\lambda_A}/\mu^+}\ \otimes\
    f_{W^-_{\lambda_B}/\mu^-}\ 
    \nonumber\\ 
    &\ 
    \otimes\
    \hat{\sigma}_{W^+W^-\to hh}(\lambda_A,\lambda_B)
    \\
    &= \int_{\tau_{\rm min}}^1 d\tau\ \int_{\tau}^1 \frac{dz_1}{z_1}\ 
    f_{W^+_{\lambda_A}/\mu^+}(z_1,\mu_f)\ \times
    \nonumber\\
    f_{W^-_{\lambda_B}/\mu^-}&(z_2,\mu_f)\ \times
    \hat{\sigma}_{W^+W^-\to hh}(\lambda_A,\lambda_B)\ ,
    \\
    & \text{with}\
    z_2 = \frac{\tau}{z_1}\ \text{and}\ \tau_{\rm min} = \frac{(2m_h)^2}{s}\ .
\end{align}
Importantly, we impose the additional threshold condition that 
$f_{W/\mu}(z_i)=0$ for $E_{W_i} = z_i \sqrt{s}/2 < M_W$.

Unpolarized cross sections are recovered through 
averaging over initial-state spin configurations:
\begin{align}
    \hat{\sigma}_{\rm unpol} &= \frac{1}{3^2}\ \sum_{\lambda_A,\lambda_B\in\{\pm1,0\}}\ \hat{\sigma}(\lambda_A,\lambda_B)\ , 
    \\
    \sigma &= \frac{1}{3^2}\ \sum_{\lambda_A,\lambda_B\in\{\pm1,0\}}\
    \sigma_{\lambda_A\lambda_B}\ .
\end{align}
We summarize our results in Table~\ref{tab:app_checks}
for the following SM and collider inputs
\begin{align}
\label{eq:app_inputs}
    g &\approx 0.6532,\ 
    M_W \approx 80.42\GeV,\ 
    m_h = 125\GeV,
    \nonumber\\
    M_Z &\approx 91.19\GeV,\ 
    \sqrt{s} = 5\TeV,\ 
    \mu_f = m_h\ .
\end{align}

\section{Scale variation at LP and NLP}
\label{app:scale}

Assuming a collection of events, 
each with cross section weight $\Delta \sigma_k$
at a baseline scale $\mu=\mu_0$, then the scale variation 
of the cross section weight $(\Delta \sigma_k')$ at a scale 
$\mu=\zeta \mu_0$ is given by the reweighting scheme
\begin{align}
    \Delta \sigma_k'(\zeta\mu_0)\ =\ \Delta \sigma_k(\mu_0)\ \times\ \frac{w_k(\zeta\mu_0)}{w_k(\mu_0)}\  .
\end{align}
Here, $w_k(\mu)$ is the scale weight. The scale weight varies according to PDF species and accuracy.

For transverse PDFs at LLA, LP [Eqs.~\eqref{eq:pdf_xlp_vp}-\eqref{eq:pdf_xlp_vm}], and NLP [Eqs.~\eqref{eq:pdf_nlp_vp}-\eqref{eq:pdf_nlp_vm}], the scale weights are
\begin{align}
    w_k^{\rm LLA}(V_{\lambda=\pm},\mu) &=  \log\left(\frac{\mu^2}{M_V^2}\right)\ ,\\
    w_k^{\rm LP}(V_{\lambda=\pm},\mu)  &= \log\left(\frac{M_V^2+\mu^2}{M_V^2}\right)\nonumber\\
    & -\left(\frac{\mu^2}{M_V^2 + \mu^2}\right)\ ,
    \\
    w_k^{\rm NLP}(V_{\lambda=\pm},\mu) &=     
    (a_\lambda+ b_T)\log\left(\frac{M_V^2+\mu^2}{M_V^2}\right) 
    \nonumber\\ 
    & - \left(\frac{\mu^2}{\mu^2+M_V^2}\right)(a_\lambda + c_T)\ ,
\end{align}
with coefficients 
\begin{align}
    a_{\lambda=+} &= (1-z), \quad 
    a_{\lambda=-} = 1, \nonumber\\
    b_T&=(2-z)\frac{M_V^2}{E_V^2}, \\
    c_T&= b_T + (2-z)\frac{\mu^2}{2E_V^2}  \ .
\end{align}

For longitudinal PDFs at LLA, LP [Eq.~\eqref{eq:pdf_xlp_v0}], and NLP [Eq.~\eqref{eq:pdf_nlp_v0}], the scale weights are respectively,
\begin{align}
    w_k^{\rm LLA}(V_{\lambda=0},\mu) &= 1\ , \\
    w_k^{\rm LP}(V_{\lambda=0},\mu)  &= 
    \left(\frac{\mu^2}{M_V^2 + \mu^2}\right)\ ,
    \\
    w_k^{\rm NLP}(V_{\lambda=0},\mu) &= \left(\frac{\mu^2}{M_V^2 + \mu^2}\right) 
    \nonumber\\
     & - \Bigg\{\left[\frac{1-(1-z)^2}{(1-z)}\right]\left(\frac{M_V^2}{4E_V^2}\right)
     \nonumber\\
      \times
     \Big[\log\left(\frac{M_V^2+\mu^2}{M_V^2}\right)
     &-\left(\frac{\mu^2}{M_V^2 + \mu^2}\right)\Big]\Bigg\}\ .
\end{align}


\bibliography{ewa_nlp_refs.bib}

\end{document}